\documentclass[journal=prb, layout=onecolumn, preprint, manuscript=article]{revtex4-2}

\usepackage{graphicx}
\usepackage{tabularx}
\usepackage{float}
\usepackage{dcolumn}
\usepackage{bm}
\usepackage{units}
\usepackage{amsmath}
\usepackage[version=4]{mhchem}
\usepackage{soul}
\usepackage[dvipsnames]{xcolor}
\usepackage{makecell}
\usepackage{listings}

\def\cm{\,\textrm{cm}}
\def\nm{\,\textrm{nm}}
\def\ang{\,\textrm{\AA}}
\def\eV{\,\textrm{eV}}
\def\meV{\,\textrm{meV}}
\def\Ry{\,\textrm{Ry}}
\def\Bohr{\,\textrm{Bohr}}
\def\K{\,\textrm{K}}

\def\eVAng{\,\textrm{eV}/\textrm{\AA}}
\def\Vnm{\,\textrm{V}/\textrm{nm}}

\def\first{1\textsuperscript{st}}
\def\mg{\ce{Mn12} | \textrm{Gr}}
\def\gga{\textrm{Gr} | \textrm{GaAs}}
\def\mgga{\ce{Mn12} | \textrm{Gr} | \textrm{GaAs}}
\def\gmgga{\textrm{Gr} | \ce{Mn12} | \textrm{Gr} | \textrm{GaAs}}
\def\gmg{\textrm{Gr} | \ce{Mn12} | \textrm{Gr}}

\begin{document}

\title{Single-Molecule Magnet Mn$_{12}$ on GaAs-supported Graphene: \\
Gate Field Effects from First Principles}

%superscript affiliations
\author{Shuanglong Liu,$^{1,2,3}$
%\email{shlufl@ufl.edu}
Maher Yazback,$^{1,2,3}$
%\email{myazback@ufl.edu}
James N. Fry,$^1$ \\
%\email{fry@ufl.edu}
Xiao-Guang Zhang,$^{1,2,3}$
% \email{xgz@ufl.edu}
Hai-Ping Cheng$^{1,2,3}$}
\email{hping@ufl.edu}

\affiliation{
$^1$Department of Physics, University of Florida, Gainesville, Florida 32611, USA \\
$^2$Quantum Theory Project, University of Florida, Gainesville, Florida 32611, USA \\
$^3$Center for Molecular Magnetic Quantum Materials, University of Florida,  Gainesville, Florida 32611, USA}

\begin{abstract}

We study gate field effects on the heterostructure 
\ce{Mn12O12(COOH)16(H2O)4} $|$ graphene $|$ GaAs via first-principles calculations. 
We find that under moderate doping levels electrons can be added to but not taken from the single-molecule
magnet \ce{Mn12O12(COOH)16(H2O)4} (\ce{Mn12}). 
%
% The magnetic anisotropy energy (MAE) of \ce{Mn12} decreases as the electron doping level increases, due to a combination of electron transfer from graphene to \ce{Mn12} and nonuniform local electric field from the substrate. 
%
The magnetic anisotropy energy (MAE) of \ce{Mn12} decreases as the electron doping level increases, due to electron transfer from graphene to \ce{Mn12} and change in the band alignment between \ce{Mn12} and graphene.
At an electron doping level of 
$-5.00 \times 10^{13} \cm{}^{-2}$, the MAE
decreases by about 18\% compared with zero doping. 
The band alignment between graphene and GaAs is more sensitive to electron
doping than to hole doping, since the valence band of GaAs is close to the Fermi level. 
The GaAs substrate induces a small band gap in the supported graphene under zero gate field and a nearly strain-free configuration. 
Finally, we propose a vertical tunnel junction for probing the
gate dependence of MAE via electron transport measurements. 

\end{abstract}

\maketitle

\section{Introduction}

\ce{Mn12O12(COOR)16(H2O)4}, where
\ce{R} represents \ce{-CH3} or other ligands, is a prototypical single-molecule magnet (SMM)~\cite{RN606}
whose magnetic and electronic properties have been studied since the late 90s~\cite{RN2490, RN1551,RN2196, RN1561, RN612}.
This molecule is also interesting as a spin system
because its total magnetic moment puts it near the boundary between classical and quantum regimes.
Tunneling magnetism measurements show its quantum nature \cite{RN3280}, but its big magnetic moment ($S=10$) makes it almost classical.
The magnetic anisotropy energy (MAE) of
\ce{Mn12O12(COOR)16(H2O)4}, 
which forms the barrier for magnetic tunneling, 
depends on the type of ligand~\cite{RN649, 
RN639, RN638, RN634, RN601, RN660, RN657, RN664, RN633, RN646, RN642, RN625, RN624, RN628, RN626, RN617, RN658, RN618, RN653, RN614, RN623, RN616, RN613, RN636, RN615}
as well as the charge state of the molecule. 
In experiments, one or two electrons reduced \ce{Mn12O12(COOR)16(H2O)4}
have been synthesized by adding \ce{PPh4^+} or other cations to the
molecular crystal~\cite{RN635, RN605, RN604,RN663, RN660, RN637}.
The resulting negatively charged \ce{Mn12O12(COOR)16(H2O)4} with integer number of electrons yields a decreased MAE for
\ce{-CHCl2}~\cite{RN604} or \ce{-C6F5}~\cite{RN660} ligands. 
Based on computations, a decreased MAE has also been reported for negatively
charged \ce{Mn12O12(COOR)16(H2O)4} with \ce{-H}~\cite{RN612} or \ce{-C6H5}~\cite{RN593} ligands. 
%
% We are not sure about the ligand for this citation: but an increased MAE for \ce{-CH3} ligands~\cite{RN1561}. 
%
When a magnetic molecule is adsorbed on a surface, a fractional number of 
electrons may be transferred to/from the molecule, modifying its MAE~\cite{RN34, RN443, RN422, RN682}.

Most earlier experiments are performed either in solution or in molecular crystals.
Recently, Hebard's~\cite{RN331} group experimentally
investigated the gate-voltage dependence of transport properties of
\ce{Mn12O12(COOR)16(H2O)4} on graphene surfaces (supported by Si) with different ligands. 
The applied gate voltage is believed to affect transport properties by
modulating the charge transfer between graphene and \ce{Mn12O12(COOR)16(H2O)4}. 
However, such experiments were unable to determine whether the MAE of
\ce{Mn12O12(COOR)16(H2O)4} is also tuned by the gate voltage and
the charge transfer induced by it.
It is also not clear how the coupling to a semiconductor substrate changes
the magnetic and electronic properties of the molecule.
In this study, we aim to answer these questions through a first-principles calculation 
using \ce{Mn12O12(COOH)16(H2O)4} (\ce{Mn12}) as an example.

% We choose GaAs-supported graphene ($\gga$) as substrate in coordination with our efforts on mobility~\cite{RN3282} and Schottky barrier studies~\cite{RN3281}. 
%
We choose GaAs-supported graphene ($\gga$) as substrate. 
Compared to a Si substrate, GaAs provides free carriers at lower temperatures~\cite{RN681, RN686}. This allows transport measurements at lower temperatures, where the magnetic properties of SMMs are better observed.
Common stable GaAs surfaces include (100)~\cite{RN669, RN670} and (111)~\cite{RN671, RN581} with various surface reconstructions.  
The GaAs (111) surface has closest lattice match with the graphene lattice.
Munshi \textit{et al.} reported the growth of GaAs nanorods on few-layer graphene, where a GaAs(111) surface is in contact with the top graphene layer~\cite{RN672}.
Several epitaxial atomic structures of the interface between GaAs(111) and graphene
have been proposed in literature but all with relatively large strain in graphene~\cite{RN672, RN599}.
Here, we propose a different atomic structure where the strain in graphene is close to zero.
Using this structure, we simulate gate field effects on the $\gga$
interface with and without the adsorption of \ce{Mn12}. 
For brevity, we denote the heterostructure with \ce{Mn12} by $\mgga$.

The rest of the paper is organized as follows. 
We describe the computational details in Section \ref{sect:method}. 
We present the atomic structure, the electronic structure, and the
magnetic anisotropy of the $\mgga$ heterostructure in Sections
\ref{sect:atomic_struct}--\ref{sect:mae}. 
In Section \ref{sect:vert_tunneling}, we propose a vertical tunnel
junction based on the $\mgga$ heterostructure. 
Finally, we conclude in Section \ref{sect:conclusion}.

\section{Method}
\label{sect:method}

All calculations are based on density functional theory (DFT)~\cite{
RN74, RN75} as implemented in the Vienna Ab initio Simulation Package (VASP)~\cite{RN76, RN77} and the SIESTA package~\cite{RN267}. 
We use VASP to relax atomic structures with no applied 
electric field and SIESTA to calculate electronic structures in the 
presence of a gate electric field. 
The atomic structure is kept at the relaxed configuration with no electric field.

In VASP calculations, we apply an energy cutoff of $ 500 \eV $ for plane waves and projector augmented wave (PAW) pseudopotentials~\cite{RN79}.
We adopt the exchange correlation energy functional proposed by Klimes \textit{et al.}, optB86b~\cite{RN675}, to include the van
der Waals interaction between graphene and \ce{Mn12} (or GaAs). 
Given the large supercell size ($a=b\approx3.2 \nm $, $ c=5.2 \nm $), 
only the $\Gamma$ point is sampled in reciprocal space.
The energy tolerance for electronic self-consistency and the force
tolerance for ionic relaxation are set to $1 \times 10^{-6} \eV $
and $0.02 \eVAng $ respectively. 
We use Gaussian smearing with a smearing parameter of $ 0.2 \eV $ 
to facilitate the electronic self-consistent procedure. 
%
% A slab of GaAs was constructed for modeling the heterostructure $\mgga$. 
%
In order to eliminate interaction between periodic images in the
perpendicular direction (with respect to the GaAs slab), a vacuum layer of
at least $ 18 \ang $ is added and electric dipole corrections
(for both energy and force) are enabled. 
On-site Coulomb interaction ($U$) within the DFT+$U$ method and
spin-orbit interactions are not considered for ionic relaxations.

In SIESTA calculations, we apply double-$\zeta$ polarized (DZP) basis functions~\cite{RN267}
for Mn and O atoms and single-$\zeta$ polarized (SZP) basis functions for C, H, Ga, and As atoms. 
Such a mixed basis set allows us to describe the magnetic properties (due to Mn atoms) accurately with less computational load.
A SZP basis set is often not sufficiently accurate for structural relaxation, but it usually produces reasonably good electronic structure for a fixed atomic structure. 
The basis functions are optimized for \ce{Mn12}, graphene, and the GaAs slab separately.
Detailed specifications of the basis functions are presented in Appendix~\ref{app:basis}.
Since the optB86b functional is not available in SIESTA and the atomic structure is fixed, we use the Perdew-Burke-Ernzerhof (PBE) exchange correlation energy functional~\cite{RN78} instead.
We apply norm-conserving pseudopotentials as generated by the 
Troullier-Martins scheme~\cite{RN271} and a mesh cutoff
of $ 200 \Ry $ for real space sampling.
To accurately determine the Fermi energy, we adopt a $6 \times 6$ $k$-grid
for sampling reciprocal space~\cite{RN314} and the
$4^\textrm{th}$ order Methfessel-Paxton smearing method~\cite{RN676} with a
smearing temperature of $ 200 \K $. 
Results of convergence tests with respect to the smearing temperature and with respect to the $k$-grid are given in Tables~\ref{tab:cvg_T} and \ref{tab:cvg_k} respectively (see Appendix~\ref{sect:convergence}).
The effects of a single back gate are modeled via the effective screening medium (ESM) method~\cite{RN92}.
In the single gate configuration, the boundary condition for the back-gate (vacuum) side is constant electrostatic potential (vanishing first derivative of the potential).
Such a non-periodic boundary condition for the electrostatic potential is imposed
by the corresponding Green's function in the ESM method. 
In our simulations, the back-gate (vacuum) boundary is $15 \ang $ below (above) the bottom (the top) of the system under study. 
In order to improve the numerical results for band alignment within the
heterostructure $\mgga$, we treat the semicore Ga $d$ electrons as valence
electrons and adopt the DFT+1/2 approach for GaAs~\cite{RN677, RN678}.
A DFT+1/2 cutoff radius of $ 3.8 \Bohr $ is applied to limit the range of
self-energy potential for As $4p$ orbitals~\cite{RN93}.
The DFT+1/2 band gap of bulk GaAs is calculated to be $ 1.526 \eV $, 
in good agreement with the experimental value of 
$ 1.52 \eV $ at low temperatures~\cite{RN681}, 
as opposed to $ 0.236 \eV $ without this approach.
%
% Ga sp 0
% Ga spd 0.2355
% Ga spd pbe-half 1.5260
% Ga spd pbe-half SOC 1.4366
%
For the same purpose of improving the band alignment, we apply the DFT+$U$
method proposed by Dudarev \textit{et al.}~\cite{RN262} and set the on-site
Coulomb interaction $U$ to $ 4 \eV $ for Mn atoms.
This value for $U$ yields good agreement of the density of states of \ce{Mn12} compared with X-ray photoemission spectra (XPS) and X-ray emission spectra (XES)
measurements~\cite{RN363}.

We also use the SIESTA package to calculate the MAE of \ce{Mn12} on graphene
without the GaAs substrate (to be justified later). 
Spin-orbit interactions are included via the pseudopotentials~\cite{RN281, RN280} and evaluated in an on-site approximation~\cite{RN300}.
At high electron doping levels, the self-consistent calculation fails
to converge with the Methfessel-Paxton smearing method. 
This convergence problem is solved by using the Fermi-Dirac smearing method. 
To improve the numerical accuracy, we apply a $24\times 24$ $k$-grid
together with a smearing parameter of $0.041 \eV$ and an energy tolerance
of $1 \times 10^{-6} \eV$ for electronic self-consistency. 
We set the spin-orbit coupling strength parameter to be 1.34, such that the calculated MAE ($ 5.2 \meV $, or $ 61 \K $) of \ce{Mn12O12(COOCH3)16(H2O)4} is
close to the experimental value~\cite{RN684}.
The DFT+$U$ method is not applied for calculating the MAE.

\section{Results}
\label{sect:results}

\subsection{Atomic structure}
\label{sect:atomic_struct}

\begin{figure}[htb!]
\centering
\includegraphics[width=1.0\textwidth]{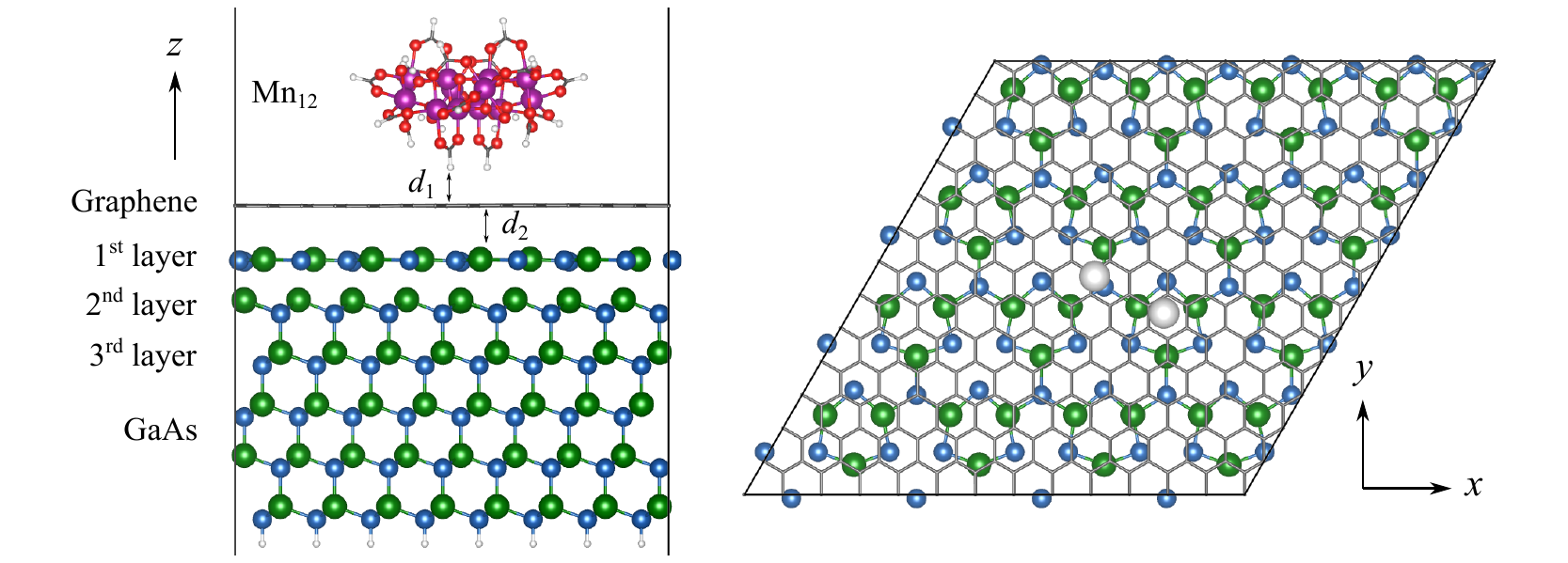}
\caption{\label{fig:pos}
Side view (left) and top view (right) of the $\mgga$ heterostructure. 
In the side view, $d_1 \approx 2.46 \ang$ 
and $d_2 \approx 3.42 \ang$. 
In the top view, only graphene and the atoms adjacent to graphene
are shown.
The hydrogen atoms of \ce{Mn12} are magnified for visibility. 
Purple: manganese, red: oxygen, gray: carbon, white: hydrogen, 
green: gallium, and blue: arsenic. 
}
\end{figure}

Fig.~\ref{fig:pos} shows the atomic structure of the heterostructure
$\mgga$. 
The GaAs(111) surface is modeled by a slab consisting of six Ga atomic
layers and six As atomic layers. 
The top two atomic layers are stabilized by a $2 \times 2$ reconstruction
with Ga vacancies~\cite{RN581}.
Each As atom at the bottom is terminated by a pseudo hydrogen atom with
0.75 electrons to avoid fictitious surface bands. 
The lattice
constants of bulk GaAs and graphene are 5.653 and $2.461 {\ang}$, respectively.
We found a good lattice match for the supercell shown in Fig.~\ref{fig:pos}, which contains $4 \times 4$ GaAs unit cells  (with surface reconstruction) 
and $13\times 13$ graphene unit cells. 
This allows graphene to match
the GaAs lattice with only a 0.04\% compression. 

\begin{figure}[H]
\centering
\includegraphics[width=0.5\textwidth]{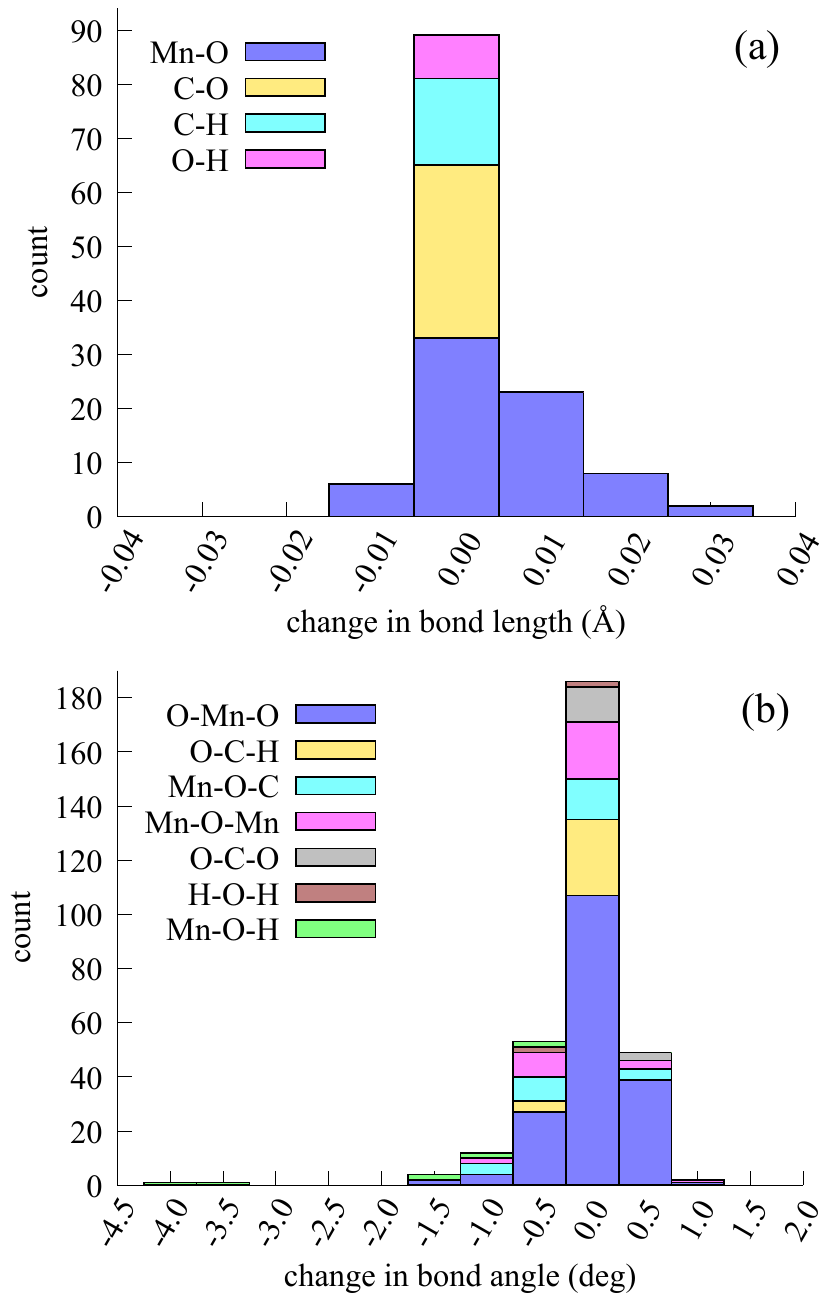}
\caption{\label{fig:hist}
Statistics of (a) bond length changes and (b) bond angle changes of \ce{Mn12}.
The comparison is between the \ce{Mn12} molecule adsorbed on $\gga$ and an isolated free \ce{Mn12} molecule.}
\end{figure}

During atomic relaxation, all atoms are relaxed except the bottom three
Ga (As) atomic layers in order to mimic the bulk environment of GaAs. 
Without the \ce{Mn12} molecule, graphene already buckles slightly due to
the nonuniform interaction with GaAs. 
The maximal out-of-plane displacement of graphene (carbon atoms) is about $\pm 0.08 \ang$,
which is one order of magnitude larger than in-plane displacements. 
After the adsorption of \ce{Mn12}, graphene is further distorted,  with a maximal out-of-plane displacement of about $\pm 0.03 \ang{}$. 
%
% The carbon atoms in contact with the \ce{Mn12} molecule move towards the GaAs substrate. 
%
Due to relatively weak chemical bonds between the core of the \ce{Mn12} molecule and the surrounding ligands, the molecule is prone to distortion and losing parts of ligands. 
For example, \ce{Mn12} loses its structural integrity when deposited on a Au(111) surface~\cite{SRN928}.
In contrast, previous DFT calculations suggest that \ce{Mn12} remains intact on graphene~\cite{RN2196}.
In the current study, the structure of \ce{Mn12} also remains intact with slight structural distortion when it is adsorbed on $\gga$. 
Figs.~\ref{fig:hist}a and \ref{fig:hist}b show histograms of the number of
chemical bond lengths and bond angles versus change in the bond length or bond angle.
All bond lengths or bond angles of \ce{Mn12} change by no more than $0.03 \ang$ or $4^\circ$.
A fraction 69.5\% of all chemical bond lengths and 60.4\% of bond angles of \ce{Mn12} change within the range of $[-0.005: 0.005] \ang{}$ and $[-0.25:0.25]\, \deg$. 
However, there are 8 \ce{Mn}-{O} bonds which change by more than
$0.015 \ang$ in length. 
Each of these \ce{Mn}-{O} bonds stems from a \ce{Mn^{3+}} ion
of the outer \ce{Mn8O8} ring, and all of them point towards graphene
(from Mn to O).
Also, there are two \ce{Mn}-{O-H} bond angles that change by more than $3.5\, \deg$. 
For both of these \ce{Mn}-{O-H} entities, the \ce{O}-{H} bond
belongs to a \ce{H_2O} unit and the \ce{Mn}-{O} bond length changes by
more than $0.015 \ang$. 

\subsection{Electronic structure}
\label{sect:elect_struct}

Now, we turn to the electronic structure of the $\mgga$ heterostructure
under both zero and finite gate electric fields. 
Without spin-orbit coupling, isolated graphene is a semi-metal, without an energy gap at the Fermi level. 
When graphene is supported on the GaAs(111) surface, calculation using SIESTA shows an energy gap of $ 2.2 \meV $ at the Dirac point. 
There are two possible factors that can induce such an energy gap in graphene,
1) a structural distortion in graphene itself, and 
2) the non-uniform potential due to the GaAs substrate.  
To identify which factor is responsible for the gap, we remove the GaAs substrate and compute the band structure for isolated graphene but with the same structural distortion.
The resulting band structure is gapless to within our numerical precision
($\sim 0.1 \meV $). 
From this we conclude that the $ 2.2 \meV $ energy gap is due to the non-uniform potential of the GaAs substrate. 
The band gap of graphene does not change after adsorption of \ce{Mn12}. 
%
% {\color{blue} Using a 6x6 (9x9) $k$-grid, the band gap of $\mgga$ is
% calculated to be 2.1 (2.2) meV.}

\begin{figure}[H]
\centering
\includegraphics[width=0.6\textwidth]{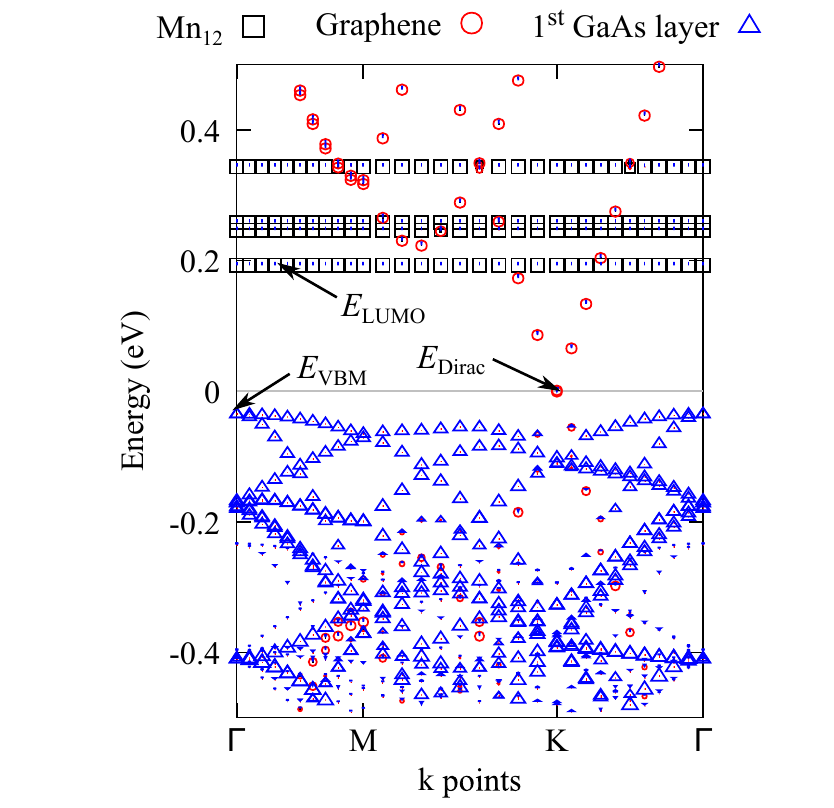}
\caption{\label{fig:fatbands}
Band structure of $\mgga$. 
Squares, circles, and triangles represent, respectively, \ce{Mn12}, graphene, and GaAs states. 
The size of a symbol (square, circle, triangle) is proportional to the
projected density of states of the \ce{Mn12} molecule,  graphene layer, or \first{} GaAs layer). 
%
% $E_\textrm{LUMO}$ is the energy of the lowest unoccupied molecular orbital
% of \ce{Mn12}. 
%
% $E_\textrm{VBM}$ is the energy of the valence band maximum (VBM) of GaAs. 
%
% $E_\textrm{Dirac}$ is the energy of the Dirac point of graphene. 
%
The Fermi level is set to zero. 
}
\end{figure}

In the $\mgga$ heterostructure, both graphene and GaAs are nonmagnetic, and according to our spin-polarized DFT calculations, 
both the highest occupied molecular orbital (HOMO) 
and the lowest unoccupied molecular orbital (LUMO) of \ce{Mn12} have the same spin. 
We define this spin to be spin up.
Fig.~\ref{fig:fatbands} shows the spin-up energy bands of this heterostructure.
The Fermi level lies within the $ 2.2 \meV $ energy gap of graphene. 
Relative to the Fermi energy, the HOMO orbital of \ce{Mn12}, the LUMO orbital of \ce{Mn12}, the valence band maximum (VBM) of GaAs and the conduction band minimum (CBM) of GaAs lie at 
$E_\textrm{HOMO} = -0.681 \eV $, 
$E_\textrm{LUMO} = 0.192 \eV$, 
$E_\textrm{VBM} = -0.035 \eV $, and 
$E_\textrm{CBM} = 1.496 \eV $. 
Compared with these four typical energies, the band gap of graphene is much smaller. 
For this reason, and for convenience of discussion,
% of discussing the band alignment between \ce{Mn12}, graphene, and GaAs,
we will denote the \textit{Dirac point} as the middle of the apexes of the upper and lower Dirac cones, even though the two cones are not quite connected. 
The LUMO energy $E_\textrm{LUMO}$, 
the energy of the Dirac point $E_\textrm{Dirac}$, 
and the valence band maximum $E_\textrm{VBM}$ 
together dictate the band alignment between \ce{Mn12},
graphene, and GaAs. 
%
% They are labeled in Fig. \ref{fig:fatbands}. 

\begin{figure}[H]
\centering
\includegraphics[width=0.5\textwidth]{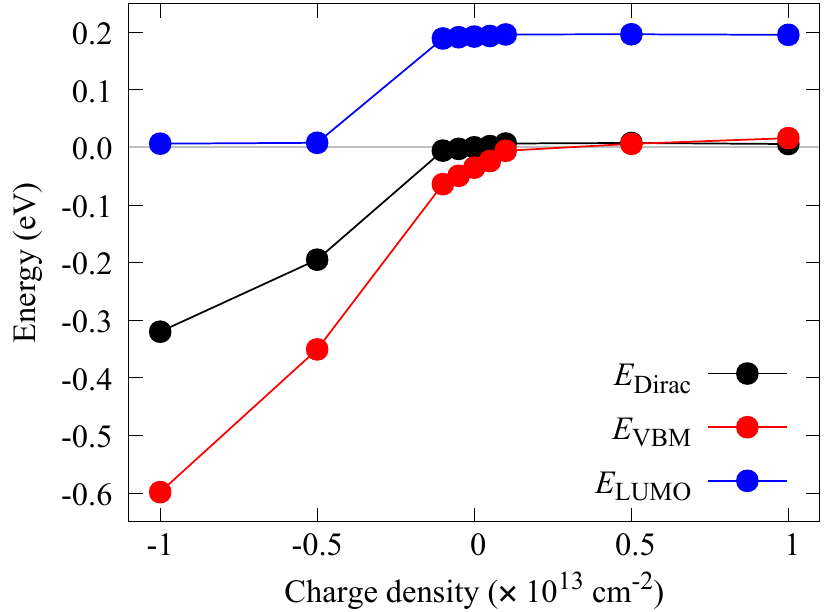}
\caption{\label{fig:typical_energies}
The typical energies $E_\textrm{LUMO}$, $E_\textrm{Dirac}$, and 
$E_\textrm{VBM}$ of $\mgga$ versus charge doping level. 
}
\end{figure}

% Next, we examine how the band alignment between \ce{Mn12}, graphene, and GaAs
% is affected by electrostatic (or charge) doping. 
%
Fig.~\ref{fig:typical_energies} shows how 
$E_\textrm{LUMO}$, $E_\textrm{VBM}$, and $E_\textrm{Dirac}$  
are affected by carrier density, i.e., the dependence of the band
alignment on electrostatic doping. 
A negative (positive) carrier density means electron (hole) doping. 
All energies are measured relative to the Fermi energy, which is set at zero. 
Overall, all these three typical energies decrease with 
electron doping and increase with hole doping level. 
$E_\textrm{Dirac}$ in particular is more sensitive to electron doping than hole doping.
For example, $E_\textrm{Dirac}$ is $-0.195 \eV$ at a charge density of $\rho = -0.50 \times 10^{13} \cm^{-2}$, but only $+0.008 \eV$ at $+0.50 \times 10^{13} \cm^{-2}$. 
The asymmetric response of $E_\textrm{Dirac}$ to charge doping can be 
understood from two aspects. 
First, electrons are mainly added to (taken from) graphene before the
HOMO of \ce{Mn12} (the VBM of GaAs) is brought to the Fermi level, whence \ce{Mn12} (GaAs) becomes charged (doped). 
However, the VBM of GaAs is much closer to the Fermi level than the HOMO of \ce{Mn12}. 
Second, GaAs has a much higher density of states than graphene. 
Therefore, the same amount of charge causes a smaller shift in the
energy bands when GaAs is doped than when graphene is doped. 
%
%The energy difference $E_\textrm{LUMO} - E_\textrm{Dirac}$ 
%($E_\textrm{Dirac} - E_\textrm{VBM}$) indicates the potential buildup
% between graphene and \ce{Mn12} (GaAs), which we will discuss later. 
%
At $\rho = -1.00 \times 10^{13} \cm^{-2}$, the LUMO orbital of \ce{Mn12} is
only a few meV above the Fermi level. 
According to our calculations, the LUMO of \ce{Mn12} becomes partially occupied,
and thus \ce{Mn12} is negatively charged. 
It is noteworthy that the Dirac point of graphene becomes lower in energy
than the VBM of GaAs at a point between 
$\rho = 0.50 \times 10^{13} \cm^{-2}$ 
and $\rho = 1.00 \times 10^{13} \cm^{-2}$. 
We also simulate gate field effects on the band alignment of $\gga$
without \ce{Mn12}. 
By comparing with $\mgga$, we find that the influence of \ce{Mn12} on both
$E_\textrm{Dirac}$ and $E_\textrm{VBM}$ is smaller than 
$ 1 \meV $ within
the charge density range of $ [-0.50:1.00] \times 10^{13} \cm^{-2}$. 
At $\rho = -1.00 \times 10^{13} \cm^{-2}$, $E_\textrm{Dirac}$
($E_\textrm{VBM}$) increases by 3.3 (2.0) meV upon the adsorption of \ce{Mn12}.

\begin{table}[H]
\caption{\label{tab:mulliken}
Mulliken charge analysis for $\mgga$ at different charge doping levels. 
Each entry represents the amount of excess charge per atom. 
Positive (negative) values indicate gain (loss) of electrons relative to the case of zero doping.
The number of atoms for \ce{Mn12}, graphene, the \first{} Ga layer, and the \first{} As layer are 100, 338, 48, and 64 respectively.
}
\begin{ruledtabular}
\begin{tabular}{rrrrrrrrr}
\thead{Doping level\\($\times 10^{13}\; \textrm{cm}^{-2}$)} & 
\thead{\ce{Mn12} \\ $(\times 10^{-3})$} & 
\thead{Graphene \\ $(\times 10^{-3})$} & 
\thead{1\textsuperscript{st} \\ Ga layer \\ $(\times 10^{-3})$} & 
\thead{1\textsuperscript{st} \\ As layer \\ $(\times 10^{-3})$} & 
\thead{2\textsuperscript{nd} \\ Ga layer \\ $(\times 10^{-3})$} & 
\thead{2\textsuperscript{nd} \\ As layer \\ $(\times 10^{-3})$} & 
\thead{3\textsuperscript{rd} \\ Ga layer \\ $(\times 10^{-3})$} & 
\thead{3\textsuperscript{rd} \\ As layer \\ $(\times 10^{-3})$} \\
\colrule
$-1.00$ &  0.61  &  1.67  & $-6.36$  & $-3.38$  &  0.26  &  0.03  &  0.17  & $-0.14$ \\
$-0.50$ &  0.00  &  0.91  & $-3.15$  & $-1.66$  &  0.13  &  0.02  &  0.08  & $-0.06$ \\
$-0.10$ &  0.00  &  0.17  & $-0.59$  & $-0.31$  &  0.03  &  0.01  &  0.01  & $-0.01$ \\
$-0.05$ &  0.00  &  0.09  & $-0.30$  & $-0.15$  &  0.01  &  0.00  &  0.01  & $-0.01$ \\
 0.00 &  0.00  &  0.00  &  0.00  &  0.00  &  0.00  &  0.00  &  0.00  &  0.00 \\
 0.05 &  0.00  & $-0.07$  &  0.25  &  0.10  &  0.00  & $-0.01$  & $-0.01$  &  0.00 \\
 0.10 &  0.00  & $-0.18$  &  0.62  &  0.30  & $-0.02$  & $-0.01$  & $-0.01$  &  0.01 \\
 0.50 &  0.00  & $-0.19$  &  0.56  &  0.13  & $-0.13$  & $-0.20$  & $-0.06$  & -0.07 \\
 1.00 &  0.00  & $-0.17$  &  0.41  & $-0.03$  & $-0.21$  & -0.36  & $-0.12$  & -0.20 \\
\end{tabular}
\end{ruledtabular}
\end{table}

Table~\ref{tab:mulliken} shows the distribution of excess charges at the
interface of $\mgga$ based on Mulliken charge analysis. 
The electron distribution for $\mgga$ under zero doping level is subtracted as a reference. 
As seen from the table, the number of electrons on \ce{Mn12} barely changes
except at $\rho = -1.00 \times 10^{13} \cm^{-2}$, 
where 0.06 ($\approx 0.61 \times 10^{-3} \times 100$) electrons are added to \ce{Mn12}. 
Graphene gains 0.56 ($\approx 1.67 \times 10^{-3} \times 338$) 
at a doping level of $\rho = -1.00 \times 10^{13} \cm^{-2}$, and
it loses 0.06 ($\approx \lvert -0.17 \rvert \times 10^{-3} \times 338$) electrons
at a doping level of  $+1.00 \times 10^{13} \cm^{-2}$.
This is consistent with our previous observation that electron doping results
in a larger shift of the Dirac point than hole doping. 
The \first{} GaAs layer tends to lose (gain) electrons when graphene gains
(loses) electrons. 
This is likely due to the Coulomb repulsion between these two adjacent
atomic layers. 
The number of electrons gained by the \first{} GaAs layer does not change
monotonically with the hole doping level, but reaches a maximum at $\rho = +0.10 \times 10^{13} \cm^{-2}$. 
This observation supports that GaAs is doped with holes 
at a doping level between 
$\rho = +0.10 \times 10^{13} \cm^{-2}$ and 
$\rho = +0.50 \times 10^{13} \cm^{-2}$.

\begin{figure}[htb!]
\centering
\includegraphics[width=0.5\textwidth]{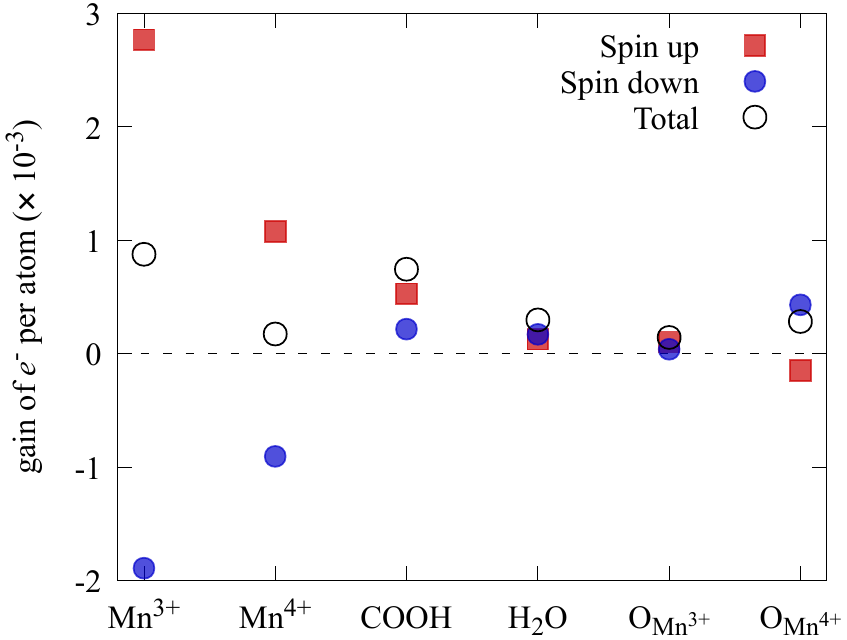}
\caption{\label{fig:mulliken_mn12}
Mulliken charge analysis for \ce{Mn12} within $\mgga$ at a doping level of $\rho = -1.00 \times 10^{13} \cm^{-2}$. 
The charge distribution at zero doping is subtracted as a reference. 
$\textrm{O}_{\textrm{Mn}^\textrm{3+}}$ ($\textrm{O}_{\textrm{Mn}^\textrm{4+}}$) denotes oxygen atoms in the \ce{Mn8O8} ring (\ce{Mn4O4} center).}
\end{figure}

% In the following, we present electron redistribution in the \ce{Mn12} molecule at the charge doping level of $\rho = -1.00 \times 10^{13} \cm^{-2}$ based on Mulliken charge analysis. 
%
Compared with zero doping, the \ce{Mn8O8} ring, the \ce{Mn4O4} core, the \ce{COOH} groups, and the \ce{H2O} units of \ce{Mn12} gain $8.18 \times 10^{-3}$, $1.84 \times 10^{-3}$, $47.70 \times 10^{-3}$, and $3.59 \times 10^{-3}$ electrons respectively at the charge doping level of $\rho = -1.00 \times 10^{13} \cm^{-2}$. 
Although the \ce{COOH} groups gain much more charge than the \ce{Mn8O8} ring in total, one \ce{Mn^{3+}} ion (in the \ce{Mn8O8} ring) gains more electrons than one atom in the \ce{COOH} groups on average (see Fig.~\ref{fig:mulliken_mn12}).
The average amount of electrons gained by one \ce{Mn^{3+}} ion % $0.88 \times 10^{-3}$ electrons
is about four times larger than that gained by one \ce{Mn^{4+}} ion (in the \ce{Mn4O4} core). 
For both \ce{Mn^{3+}} and \ce{Mn^{4+}} ions, the spin-up channel gains and the spin-down channel loses electrons.
Since the highest occupied spin down molecular orbital is well below the Fermi level and thus remains fully occupied, the loss of electrons in the spin down channel is due to the deformation of occupied states. 
According to Boukhvalov \textit{et al.}~\cite{SRN586}, redistribution of electron density affects the magnetic exchange interactions between Mn atoms.

\begin{figure}[htb!]
\centering
\includegraphics[width=0.5\textwidth]{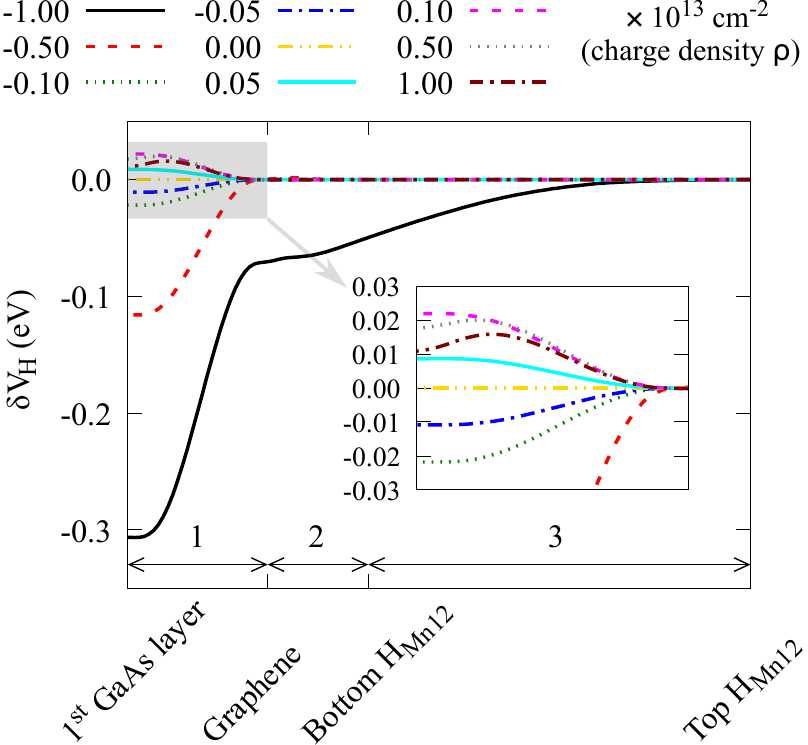}
\caption{\label{fig:dvh}
Hartree potential difference $\delta V_\textrm{H} = V_\textrm{H}(\rho, z) - V_\textrm{H}(0, z)$ of $\mgga$. 
The horizontal arrows indicate the three regions indexed by 1, 2, and 3.
\ce{H_{Mn12}} denotes a \ce{-H} ligand of \ce{Mn12}. 
The inset zooms in on the region shaded in gray.
}
\end{figure}

% Now, we examine how the Hartree potential of $\mgga$ is affected by
% electrostatic doping. 
%
Let $V_\textrm{H}(\rho, z)$ be the plane-averaged 
(over the $x$-$y$ plane) Hartree potential of $\mgga$ at position $z$ 
for charge doping density $\rho$. 
Fig.~\ref{fig:dvh} shows $\delta V_\textrm{H} = V_\textrm{H}(\rho, z) - V_\textrm{H}(0, z)$, 
which can be viewed as the gate potential. 
We focus on three regions: 1) the region between GaAs and graphene,
2) the region between graphene and \ce{Mn12}, and 3) the region beyond \ce{Mn12}. 
Let regions $i$ ($i=1,2,$ and $3$) 
begin at $z_1^i$ and end at $z_2^i$.
We define the gate potential buildup across region $i$ to
be $\Delta_i =\delta V_\textrm{H}(z_2^i) - \delta V_\textrm{H}(z_1^i)$, 
where the argument $\rho$ is dropped for brevity. 
Within the charge density range $[-0.50:1.00] \times 10^{13} \cm^{-2}$,
$\Delta_1$ is effectively tuned, whereas $\Delta_2$ is not. 
This is consistent with Fig.~\ref{fig:typical_energies}, in the sense that 
the band alignment between graphene and GaAs (\ce{Mn12}) 
% as indicated by the energy difference $E_\textrm{Dirac} - E_\textrm{VBM}$
% ($E_\textrm{LUMO} - E_\textrm{Dirac}$) 
is (is not) effectively tuned. 
Within the same charge density range, $\Delta_3$ remains almost zero, which signifies that the gate electric field does not extend across \ce{Mn12}. 
At $\rho = -1.00 \times 10^{13} \cm^{-2}$, both $\Delta_2$ and $\Delta_3$
are significantly tuned as a result of \ce{Mn12} being doped with electrons. 
The gate electric field $\mathcal{E}_i$ in region $i$ is determined
by $e \mathcal{E}_i = \Delta_i/(z_2^i - z_1^i)$ where $e$ is the unit charge (a positive value). 
$\mathcal{E}_1$, $\mathcal{E}_2$, and $\mathcal{E}_3$ are 0.69, 0.08, and $ 0.05 \Vnm$ respectively at $\rho = -1.00 \times 10^{13} \cm^{-2}$. 
Recall that GaAs becomes doped with holes at a doping level between $\rho =
0.10 \times 10^{13} \cm^{-2}$ and $\rho = 0.50 \times 10^{13} \cm^{-2}$. 
As consequence, $\mathcal{E}_1 = -0.06 \Vnm{}$ at $\rho = 0.10 \times 10^{13}
\cm^{-2}$ is larger in magnitude than $\mathcal{E}_1 = -0.05 \Vnm{}$ at
$\rho = 0.50 \times 10^{13} \cm^{-2}$, although the doping level for the latter is higher. 
Note that the adsorption of \ce{Mn12} does not affect the Hartree potential between GaAs and graphene by much (less than $1 \meV $) within the charge
density range $[-1.00:1.00] \times 10^{13} \cm^{-2}$. 
This implies that the electron tunneling rate between GaAs and graphene in transport measurements is similar with and without the presence of \ce{Mn12}.

\subsection{Magnetic anisotropy}
\label{sect:mae}

\begin{figure}[htb!]
\centering
\includegraphics[width=0.5\textwidth]{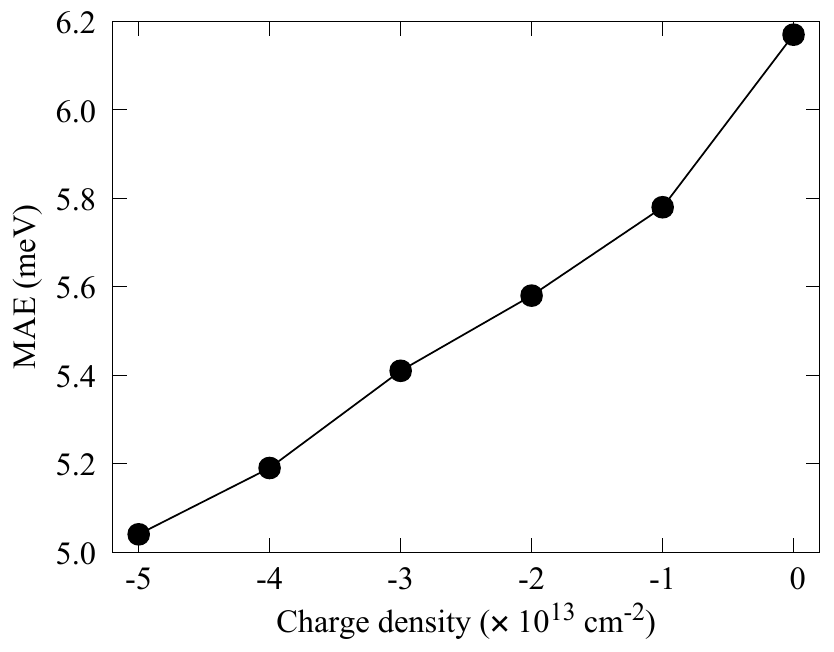}
\caption{\label{fig:mae}
Magnetic anisotropy energy (MAE) versus charge doping density for $\mg$. 
}
\end{figure}

Next, we examine how electrostatic doping affects the magnetic anisotropy energy of \ce{Mn12}. 
We have shown in Section \ref{sect:elect_struct} that electrons can be added to but cannot be taken from \ce{Mn12} at moderate doping levels. 
Therefore, we consider the effects of electron doping on MAE in particular. 
When the charge doping density is higher in magnitude than $-1.00 \times 10^{13} \cm{}^{-2}$, there are only graphene and \ce{Mn12} states near the Fermi level. 
According to second order perturbation theory~\cite{RN445}, the MAE is dominated by pairs of occupied and unoccupied states around the Fermi level. 
Therefore, it should be a good approximation to calculate MAE without GaAs. 
Based on the heterostructure $\mg$ without GaAs, we calculate the MAE as
$\mathrm{MAE} = E_{\perp} - E_{\parallel}$, where $E_{\perp}$ ($E_{\parallel}$) is the DFT total energy for the spin of \ce{Mn12} perpendicular (parallel) to the magnetic easy axis.
The calculated MAE of $\mg$ versus the charge doping density is shown in Fig.~\ref{fig:mae}. 
As seen from the figure, the MAE decreases as the electron doping level increases. 
The MAE at $\rho = -5.0 \times 10^{13} \cm{}^{-2}$ decreases by about 18\%  compared with the value at zero doping. 
It is noteworthy that the MAE of $\mg$ at zero doping is 1\% smaller than the MAE of an isolated \ce{Mn12}~\cite{RN2196}.

\begin{figure}[htb!]
\centering
\includegraphics[width=0.5\textwidth]{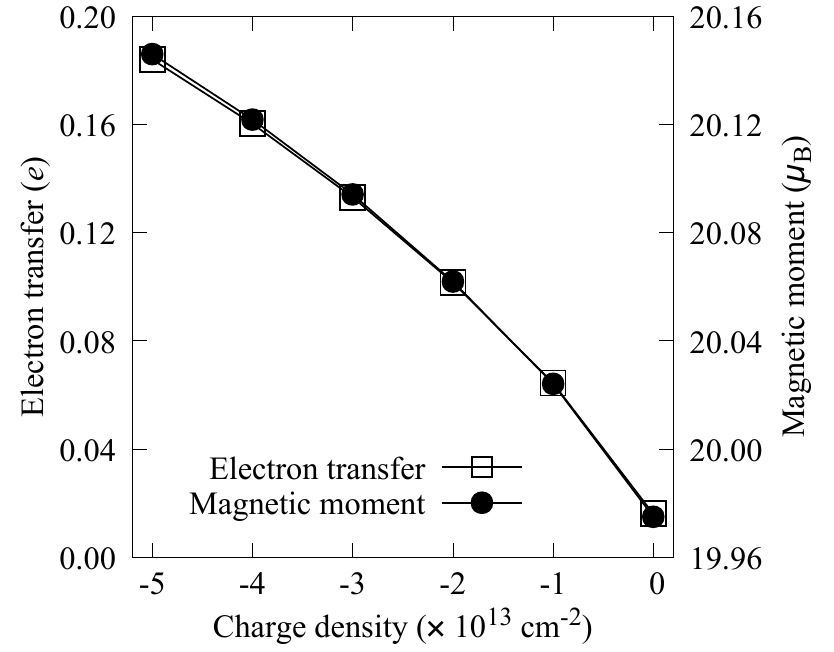}
\caption{\label{fig:charge_transfer}
Number of electrons transferred from graphene to \ce{Mn12} (left-hand scale) and magnetic moment of \ce{Mn12} (right-hand scale) versus electrostatic doping level.  
A negative charge density means electron doping. 
}
\end{figure}

We can relate the decrease in MAE in terms of electron
transfer from graphene to \ce{Mn12}. 
Fig.~\ref{fig:charge_transfer} shows the number of electrons transferred
from graphene to \ce{Mn12} versus the charge doping density. 
The number of electrons added to \ce{Mn12} increases with the electron doping
density, and about 0.18 electrons are added to \ce{Mn12} at
$\rho = -5.0 \times 10^{13} \cm{}^{-2}$. 
Along with the electron transfer, the magnetic moment of \ce{Mn12}, 
in units of Bohr magneton, increases by nearly the same amount as the number of electrons transferred.
The magnetic moment of \ce{Mn12} increases rather than decreases since the spin polarization of the LUMO orbital, which receives the added electrons, is parallel to the total spin of the whole \ce{Mn12} molecule. 
Previously, Park and Pederson considered potassium addition to introduce extra electrons on \ce{Mn12}~\cite{RN612}.
They found that the MAE of \ce{Mn12} decreases by 15\%, 37\%, and 56\% with 1, 2, and 4 extra electrons respectively. 
As a numerical experiment, we add extra electrons to a single isolated \ce{Mn12} molecule which has the same atomic positions as the \ce{Mn12} molecule of the $\mg$ heterostructure. 
The MAE of the isolated \ce{Mn12} molecule decreases as the number of added electrons increases from zero to one. 
The MAE of isolated \ce{Mn12} decreases by 15\% with one additional electron, coincident with Park and Pederson's findings. 
This confirms that electron transfer from graphene to \ce{Mn12} is responsible for the decrease in the MAE of $\mg$ under electron doping.

However, with 0.18 additional electrons, the MAE of isolated
\ce{Mn12} decreases by 2\%, which is much smaller than the decrease
of 18\% in the MAE of $\mg$. 
This can be understood within second order perturbation theory, 
where pairs of occupied
and unoccupied states can be classified into three types, 
\ce{Mn12}-\ce{Mn12}, graphene-graphene, and \ce{Mn12}-graphene. 
The 2\% decrease in the MAE of the isolated \ce{Mn12} can be
understood as a decrease in the contribution from the \ce{Mn12}-\ce{Mn12} pairs. 
Since contributions to the MAE from graphene-graphene pairs are
negligibly small (as seen from isolated graphene), 
we owe the extra reduction in the MAE of $\mg$ to a decrease in the contribution from the \ce{Mn12}-graphene pairs. 
Such an argument is supported by the observation that electron
doping dramatically changes the band alignment between \ce{Mn12} and
graphene and thus the pairs of occupied graphene (\ce{Mn12}) states
and unoccupied \ce{Mn12} (graphene) states around the Fermi level.
Note that the change in the band alignment between \ce{Mn12} and graphene
is mainly caused by electrostatic doping rather than by electron
transfer between \ce{Mn12} and graphene. 
Thus, the change in the band alignment between \ce{Mn12} and
graphene is another factor that is responsible for the decrease
in the MAE of $\mg$ under electron doping.

In order to investigate the possible role of a local electric field between
\ce{Mn12} and graphene, we take the isolated \ce{Mn12} molecule again
and move it close to ($2.46 \ang{}$ away from) the plane where the
boundary condition of constant electrostatic potential is enforced. 
In this way, we retain a strong local electric field below \ce{Mn12}
when the \ce{Mn12} molecule is charged.  
Otherwise, the local electric field is not as strong in our earlier
numerical experiment using the isolated \ce{Mn12} molecule. 
It turns out that the decrease in MAE is still 2\% with 0.18
additional electrons even though a strong local electric field is present below \ce{Mn12}. 
Therefore, we exclude the local electric field between \ce{Mn12} and
graphene and the associated electric field across \ce{Mn12} as a separate
reason for the decrease in the MAE of $\mg$ under electron doping. 
From perturbation theory~\cite{SRN698}, magnetic anisotropy of molecules is closely related to a molecular quadrupole moment, which should change with gate electric field. 
Since we determine molecular orbitals and thus electron density self-consistently, the change in molecular quadrupole moment and its influence on MAE have been captured implicitly by our calculations.
However, the effect of electric field on MAE via spin-orbit coupling amplitude is still unknown.

\subsection{A vertical tunnel junction}
\label{sect:vert_tunneling} 

\begin{figure}[htb!]
\centering
\includegraphics[width=0.9\textwidth]{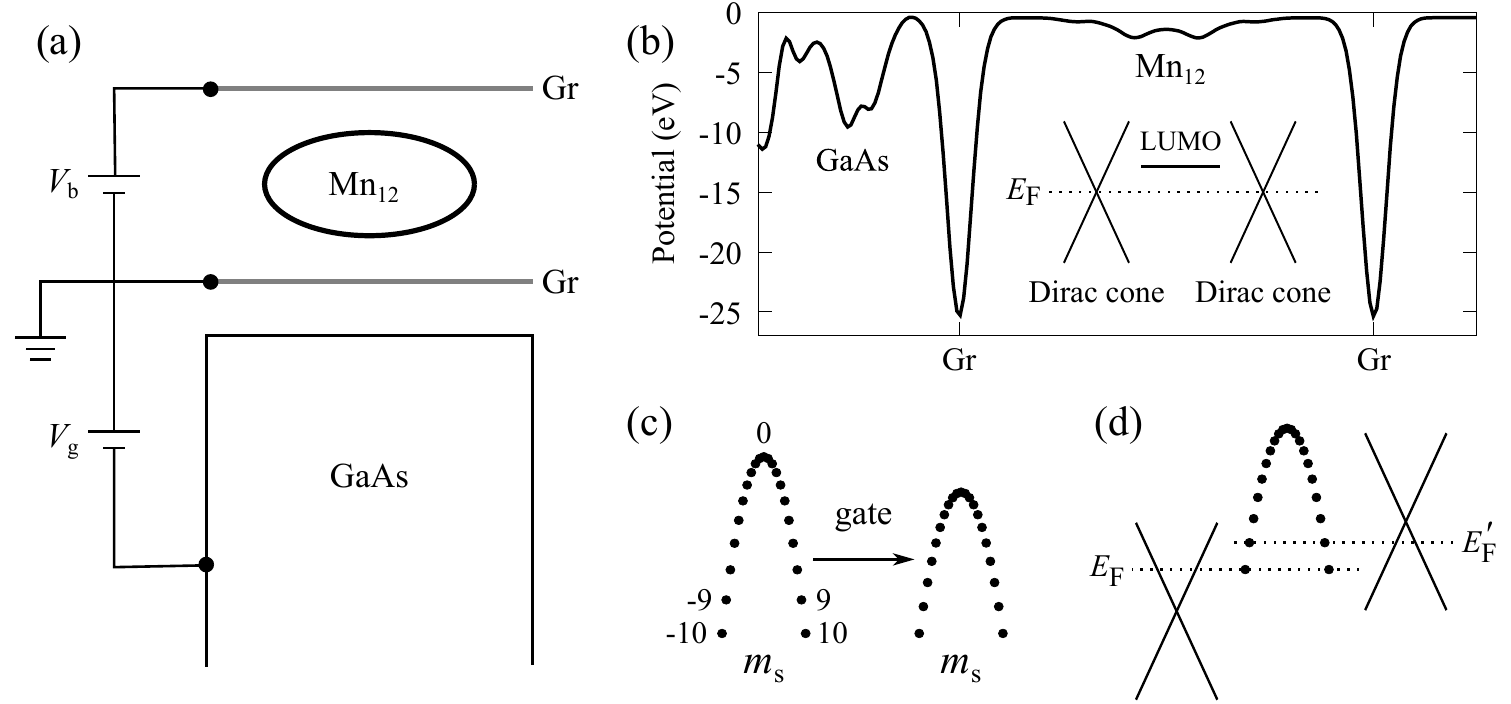}
\caption{\label{fig:grmn12gr}
(a) Illustration of a $\gmgga$ tunnel junction.
$V_g$ is gate voltage, and $V_b$ is bias voltage.
The two graphene layers are separated by 
about $ 1.42 \nm $. 
(b) Calculated plane-averaged electrostatic potential of the junction. 
The inset shows schematic band alignment between the \ce{Mn12} molecule and the two graphene layers.
(c) Spin excitations with zero gate and with a finite gate. 
$\lvert m_\textrm{s} \rangle$ are spin states
with $m_\textrm{s} = \pm 10, \, \pm 9, \, \dots, \, 0$. 
(d) Band alignment when a spin excitation enters the bias window. 
}
\end{figure}

In this section, we propose a vertical tunnel junction for detecting the
gate-dependence of MAE discussed in the previous section. 
The junction is illustrated in Fig.~\ref{fig:grmn12gr}a where a \ce{Mn12}
molecule is sandwiched between two graphene layers. 
The bottom graphene layer should be supported by $n$-type GaAs, which has
conducting charge carriers at low temperature 
(several Kelvins)~\cite{RN681, RN686}.
A gate voltage $V_\textrm{g}$is applied between the bottom graphene layer and the GaAs and a bias voltage $V_\textrm{b}$ between the bottom 
and top graphene layers.
In the proposed vertical tunnel junction, GaAs is electron-doped to bring the conduction band near the Fermi level.
Such a band alignment may change the behavior of the system substantially compared to the results presented in previous sections. 
Assuming a similar response of MAE to a gate electric field, we argue that the gate-dependence of MAE can be probed by measuring the electron tunneling current $I$ through \ce{Mn12} as a function of $V_\textrm{g}$ and $V_\textrm{b}$.

First, we construct an atomic structure of the tunnel junction by adding 
a flat graphene layer above the $\mgga$ heterostructure, and calculate from
first principles the electronic structure of the junction at zero gate and zero bias. 
%
% The distance between the top graphene layer and the \ce{Mn12} molecule was set to
% be the same as that between the bottom graphene layer and the \ce{Mn12} molecule. 
%
% The atomic structure of the top graphene layer is fixed. 
%
Fig.~\ref{fig:grmn12gr}b shows the plane-averaged electrostatic potential
(Hartree potential and ionic potential) across the tunnel junction. 
The electrostatic potential around the bottom graphene layer is similar to
that around the top graphene layer. 
The inset of Fig.~\ref{fig:grmn12gr}b depicts a schematic band alignment of the junction. 
According to our calculations, the Dirac points of either graphene layer are at the Fermi level 
and the LUMO of \ce{Mn12} is $ 0.230 \eV $ above the Fermi level. 
%
% Compared with $\mgga$, the LUMO of \ce{Mn12} is $ 0.038 \eV $$ higher in energy. 
%
To estimate the structural change in \ce{Mn12} caused by a finite bias voltage, we apply static electric fields on a single isolated \ce{Mn12} molecule.
At $0.2\, \textrm{V/\AA}$ ($\approx 2.84\, \textrm{V}$ across the junction), the maximal change in \ce{Mn}-{O} bonds is $0.032 \ang $ and the change in all other chemical bonds is below $0.003 \ang $. 
At $0.1\, \textrm{V/\AA}$, the change in all bond lengths is within $0.018 \ang $. 
Because the $S_z$ degeneracy is broken by the spin-orbit interaction,
below the LUMO energy there should be spin states of \ce{Mn12} on an 
energy scale of $ 1 \meV $ relative to the ground state energy of \ce{Mn12}. 
Such spin states are detectable via transport measurements in single-molecule
break junctions based on previous studies~\cite{RN610, RN603, RN35}.

The mechanism for detecting spin excitations in a tunneling current is similar to the approach for magnetic tunnel junctions~\cite{Zhang1997,Du2010}.
The spin excitations can also be probed in the vertical tunnel
junction shown in Fig.~\ref{fig:grmn12gr}a. 
As established in Section~\ref{sect:mae}, the MAE of \ce{Mn12} can be reduced by electron doping. 
Consequently, the energy spacings between spin excitations are reduced, as illustrated in Fig.~\ref{fig:grmn12gr}c. 
This is likely the case even though \ce{Mn12} is fractionally charged and
the spin projection $m_\textrm{s}$ is no longer a good quantum index. 
At finite bias voltages, the Fermi level of the bottom graphene layer
$E_\textrm{F}$ is different from that of the top graphene $E_\textrm{F}^\prime$.
We call the energy range between $E_\textrm{F}$ and $E_\textrm{F}^\prime$ the bias window. 
Whenever a spin excitation enters the bias window 
(see an illustration in Fig.~\ref{fig:grmn12gr}d), an additional transmission path through the excitation
of the spin state is opened, leading to a sudden increase in the tunneling current $I$. 
The record of multiple spin excitations allows determination of the energy spacings between these spin excitations, and thus the MAE of \ce{Mn12}. 
As gate voltage varies, the gate-dependence of MAE can then be probed.

In general, one can expect sensitive change of the tunneling current due to the spin state of the SMM in the junction. Therefore tunneling measurements can be used as a more general probe of the spin state of the SMM than just the MAE. 
We comment that the usual DFT plus non-equilibrium Green's function (DFT$+$NEGF) approach~\cite{SRN268} cannot capture the transport signals due to spin excitations. 
See Appendix~\ref{app:DFT_NEGF} for DFT$+$NEGF results at zero bias voltage.

\section{Conclusion and discussion}
\label{sect:conclusion}

In conclusion, we have simulated gate field effects on the heterostructures
$\mgga$, $\gga$, and $\mg$. 
In the $\mgga$ heterostructure, \ce{Mn12} can gain but not lose electrons
at moderate doping levels. 
The MAE of \ce{Mn12} adsorbed on graphene decreases by about 18\% at an electron
doping level of $-5.00 \times 10^{13} \cm{}^{-2}$. 
Such a decrease in the MAE is due to electron transfer from graphene
to \ce{Mn12} as well as the change in the band alignment between Mn12 and graphene.
%
% We proposed a $\gmgga$ vertical tunnel junction for probing the gate dependence
% of MAE via electron transport measurements. 
%
The band alignment between graphene and GaAs is more sensitive to
electron doping than to hole doping. 
At an electron doping level of $-1.00 \times 10^{13} \cm{}^{-2}$, 
the Dirac point of graphene is lifted by about $ 0.24 \eV $ relative to
the valence band maximum of GaAs. 
Compared with electrostatic doping, the adsorption of \ce{Mn12} does not 
have much effect on the band alignment between graphene and GaAs. 
The GaAs substrate induces a band gap of about $ 2.2 \meV $ in graphene due to
the interaction between GaAs and graphene. 
It remains to study gate field effects on similar heterostructures with
other ligands around the \ce{Mn12O12(COOR)16(H2O)4} molecule. 

\begin{acknowledgments}
This work was supported by the Center for Molecular Magnetic Quantum Materials, an Energy Frontier Research Center funded by the U.S. Department of Energy, Office of Science, Basic Energy Sciences under Award No. DE-SC0019330. Computations were done using the utilities of the National Energy Research Scientific Computing Center and the University of Florida Research Computing. 
\end{acknowledgments}

\appendix

\renewcommand{\thefigure}{\thesection\arabic{figure}}
\renewcommand{\thetable}{\thesection\arabic{table}}

\setcounter{figure}{0}
\setcounter{table}{0}

\section{Basis set in SIESTA calculations}
\label{app:basis} 

Our input specifications for basis set in SIESTA are given below. 
For accurate description of magnetic properties, we applied DZP basis functions for Mn and O atoms. 
The remaining atoms (C, H, Ga, and As) employ SZP basis functions. 
The basis set for C atoms of \ce{Mn12} is different from those for graphene
due to the different chemical environment, 
and the two basis sets are optimized for \ce{Mn12} and graphene separately.

\begin{center}
\begin{lstlisting}[basicstyle=\linespread{0.6}\footnotesize]
%block PAO.Basis
Mn                    2
 n=4   0   2 P   1
   6.4382730        5.1834909
   1.000      1.000
 n=3   2   2
   4.8440717        2.8316085
   1.000      1.000
O                     2
 n=2   0   2
   4.3035057        2.9518194
   1.000      1.000
 n=2   1   2 P   1
   4.9096245        3.3500338
   1.000      1.000
H                     1
 n=1   0   1 P   1
   5.2026590
   1.000
C_Mn12                2
 n=2   0   1
   4.8900415
   1.000
 n=2   1   1 P   1
   5.8351229
   1.000
C_Gr                  2
 n=2   0   1
   4.7137437
   1.000
 n=2   1   1 P   1
   4.7446991
   1.000
Ga                    3
 n=3   2   1
   4.9967221
   1.000
 n=4   0   1
   5.9850501
   1.000   
 n=4   1   1 P   1
   8.4542189
   1.000   
As                    2
 n=4   0   1
   5.0625027
   1.000   
 n=4   1   1 P   1
   5.2670527
   1.000   
H.75                  1
 n=1   0   1 P   1
   5.5154366
   1.000
%endblock PAO.Basis
\end{lstlisting}
\end{center}

\section{Convergence tests}
\label{sect:convergence}

Table~\ref{tab:cvg_T} shows test results on $\mg$ for different smearing temperatures.
When the smearing temperature is reduced from $400 \K$ to $50 \K$, the number of electrons on \ce{Mn12} (Mulliken charge) varies by less than $1\times{10}^{-5}$,  
the total magnetic moment of the hetero-structure varies by less than $0.01\ \mu_B$,  
and the position (in energy) of the Dirac point varies by less than $0.1\ \meV$. 
This indicates that a smearing temperature of $200 \K$ yields reliable electronic and magnetic properties. 
Table~\ref{tab:cvg_k} shows test results on $\mg$ for different $k$-point grids.
All $k$-point grids contain the K point where the Dirac point lies, and they yield the same energy of the Dirac point. 
The variation in both the number of electrons and the total magnetic moment are insignificant when the $k$-point grid is denser than $3\times3$.
These test results support that a $6\times6$ $k$-point grid yields reliable electronic and magnetic properties.

\begin{table}[H]
\caption{\label{tab:cvg_T}
Convergence test on $\mg$ with respect to the smearing temperature $T$. 
The $k$-point grid is fixed at $6 \times 6$. 
$N_e$ is the number of electrons on \ce{Mn12} (Mulliken charge).
$M$ is total magnetic moment. 
$E_\textrm{Dirac}$ is the position (in energy) of the Dirac point relative to the Fermi energy.}
\begin{ruledtabular}
\begin{tabular}{cccc}
$T$ (K) & 
$N_e$ & 
$M$ ($\mu_B$) & 
$E_\textrm{Dirac}$ (meV) \\
\colrule
 400 &  460.036436  &  19.9992  &  0.1 \\
 300 &  460.036438  &  19.9989  &  0.1 \\
 200 &  460.036437  &  19.9983  &  0.1 \\
 100 &  460.036434  &  19.9964  &  0.1 \\
  50 &  460.036437  &  19.9918  &  0.1 \\
\end{tabular}
\end{ruledtabular}
\end{table}

\begin{table}[H]
\caption{\label{tab:cvg_k}
Convergence test on $\mg$ with respect to the $k$-point grid.
The smearing temperature $T$ is fixed at 200 K. 
The quantities are the same with those in Table~\ref{tab:cvg_T}.}
\begin{ruledtabular}
\begin{tabular}{cccc}
$k$-grid & 
$N_e$ & 
$M$ ($\mu_B$) & 
$E_\textrm{Dirac}$ (meV) \\
\colrule
   $3 \times 3$ &  460.036385  &  19.9917  &   0.1 \\
   $6 \times 6$ &  460.036437  &  19.9983  &   0.1 \\
   $9 \times 9$ &  460.036448  &  19.9994  &   0.1 \\
 $12 \times 12$ &  460.036450  &  19.9997  &   0.1 \\
\end{tabular}
\end{ruledtabular}
\end{table}

\section{Electron Transmission} 
\label{app:DFT_NEGF} 

\begin{figure}[htb!]
\centering
\includegraphics[width=0.5\textwidth]{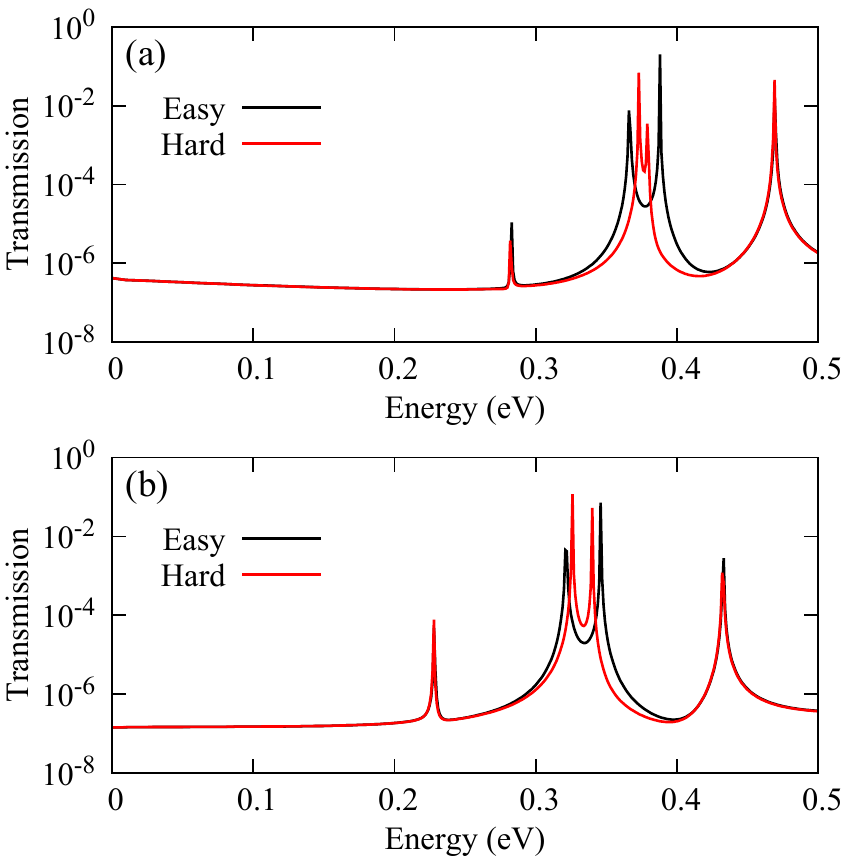}
\caption{\label{fig:transmission}
Electron transmission of a $\gmg$ junction under (a) zero doping and (b) an electron doping level of $\rho = -5 \times 10^{13}\, \textrm{cm}^{-2}$. The black (red) line is for the spin aligned in the easy (a hard) direction. The Fermi level is set to zero.  
} 
\end{figure}

We calculated electron transmission for a $\gmg$ junction under zero bias by Caroli's formula~\cite{Caroli_1971}. 
The Green's functions in Caroli's formula are based on the Kohn-Sham Hamiltonian with spin-orbit coupling. 
Due to the huge system size (1148 atoms), we included only the $\Gamma$ point for both SCF and transport calculations. 
Fig.~\ref{fig:transmission} shows the electron transmission as a function of energy above the Fermi level (set at zero). 
There are four peaks, corresponding to the LUMO (at $0.2 \eV$), LUMO+1, LUMO+2, and LUMO+3 orbitals respectively. 
There are no peaks due to spin excitations, which are a few meV above the Fermi level in Fig.~\ref{fig:transmission}. 
It requires theoretical development beyond the DFT+NEGF method to obtain electron transport signals due to spin excitations. 
It is noteworthy that both absolute and relative peak positions are tunable by electron doping as well as by spin alignment.

\bibliographystyle{apsrev4-2}
% \bibliography{references}


\end{document}